\documentclass[journal]{IEEEtran}
\usepackage{graphicx}
\usepackage{amsmath}
\usepackage{subfigure}
\usepackage{rotating}
\usepackage{multirow}
\usepackage{array}

\begin{document}

\title{Ubiquitous HealthCare in\\ Wireless Body Area Networks}

\author{\IEEEauthorblockN{N. A. Khan, N. Javaid, Z. A. Khan$^{\S}$, M. Jaffar, U. Rafiq, A. Bibi\\}

       Department of Electrical Engineering, COMSATS\\ Institute of
        Information Technology, Islamabad, Pakistan. \\
        $^{\S}$Faculty of Engineering, Dalhousie University, Halifax, Canada.
        }

\maketitle

\begin{abstract}
Recent advances in wireless communications, system on chip and low power sensor nodes allow realization of Wireless Body Area Networks (WBANs). WBANs comprise of tiny sensors, which collect information of a patient's vital signs and provide a real time feedback. In addition, WBANs also support many applications including ubiquitous healthcare, entertainment, gaming, military, etc. Ubiquitous healthcare is required by elderly people to facilitate them with instant monitoring anywhere they move around. In this paper, we provide a survey on different architectures used in WBANs for ubiquitous healthcare monitoring. Different standards and devices used in these architectures are also discussed in this paper. Finally, path loss in WBANs and its impact on communication is presented with the help of simulations performed for different models of In-Body communication and different factors (such as, attenuation, frequency, distance etc) influencing path loss in On-Body communications.
\end{abstract}


\IEEEpeerreviewmaketitle

\section{Introduction}\label{sec:intro}

\IEEEPARstart{W}{ITH} an increasing population around the world, specially the elderly people who are more fragile to health diseases, require a comprehensive healthcare system. A system fulfilling needs of elderly people provides them with proper healthcare facilities wherever, they move around. Wireless Body Area Network (WBAN) \cite{1} is gaining attention worldwide for providing healthcare infrastructure. This system consists of several devices including tiny sensors which are placed in or around the body in close proximity to monitor a patient. As a result, elderly people are monitored everywhere and treated well intime in case of any emergency. The patients specially elderly people face problems in moving around and cannot frequently visit doctor(s), indeed require Ubiquitous HealthCare (UHC) \cite{2}. Besides having applications in HealthCare, WBAN is also used in entertainment, gaming, military etc.

WBAN comprises of tiny sensors that monitor patients everywhere and reduce the number of visits to doctors. The sensors may be placed on or implanted in the body for constant monitoring. Different standards for WABN are defined which provide efficient means of data transfer and communication (such as, Bluetooth, ZigBee, MICS and UWB)\cite{3}. Similarly, different devices are used to collect patients' vital signs information and transfer it to remote healthcare personnel. These devices include wearable watch, oximeter, wearable shirt, chest belt type etc. A comprehensive and analytical survey is provided about these standards and devices in this paper. Since each architecture has its own applications, therefore, different architectures of WBAN are discussed. Several types of antennas are designed for BAN (i.e., electrical or dipole antenna and magnetic or loop antenna). Depending on the scenario of body communication (i.e., In-Body or On-Body), selection of antenna is very important, therefore, it has a direct effect on communication resulting in path loss.

Data which is collected by the sensors and devices is transferred through wireless medium to remote destination, path loss is probable to occur. Path loss for In-Body and On-Body communications are different. It depends on frequency of operations as well as distance between transmitter and receiver. A simple path loss model for WBAN is proposed in \cite{4}.

A work relating to path loss for On-body communications is provided by authors in \cite{7}. Another architecture based on traffic is discussed in \cite{8}. Where, network coordinator contains a wake up circuit to accommodate life critical events depending on traffic application. Authors also discuss other applications besides UHC. In \cite{9}, sensor devices and server based architecture for UHC monitoring system is proposed. Introduction of wireless sensor devices and server part is given in this architecture. Communication between sensor and server is done via Base Station Transceiver (BST), which is connected to a server PC.

To provide services for the elderly people, components based system architecture of UHC monitoring is designed in \cite{10}. A prototype system that monitors location and health status using Bluetooth as WBAN and smart phone with accelerometer as Intelligent Central Node (ICN) is used in this architecture. This architecture provides accessibility to family members or medical authorities to identify real time position and health status of patients via internet. ZigBee is used for small data rate applications because it consumes less power then Bluetooth.

WBAN architectures using wearable devices are proposed in \cite{11}. One of them is wearable smart shirt, which is based on UHC and activity monitoring. This architecture comprises of smart shirt with multi-hop sensor network and server PC. Communication between smart shirt and server PC is done by BST. A device with two PCB mounted on each other is used in this architecture to reduce the size of integrated wearable sensor node along with  Universal Serial Bus (USB) programming board as a separate module. It is needed only when nodes are connected to server PC.

In this paper, a comprehensive and analytical survey is provided about the standards and devices for WBAN [1-4], [7-11]. A detailed overview of UHC architectures in WBAN is provided. Also, we simulate In-Body path loss models proposed in \cite{5} using MATLAB. In simulations, we considered four path loss models; deep tissue implant to implant, near surface implant to implant, deep implant to implant, and near surface implant to implant. Further, we performed simulations on different parameters effecting (i.e., attenuation, phase distortion, RMS delay etc) communication in On-Body networks \cite{6}.

\section{Most Frequently Used Standards for WBAN Communication}\label{sec:standards}
Number of standards are adopted for communication in WBAN. Microscopic chips, which are typically used in wearable devices, depends on these standards. We briefly discuss the standards, Bluetooth, ZigBee, MICS, and Ultra Wide Band (UWB) IEEE 802.15.6 \cite{3}.

\textit{A. IEEE 802.15.1 Bluetooth:}
Bluetooth is a short range communication standard with data rate of $3 Mbps$ and range of $10m$. It is adopted in UHC due to high bandwidth and low latency. It also supports many mobile platforms. However, in UHC monitoring, use of this standard is avoided because of high power consumption. It is suitable for latency and bandwidth sensitive scenarios \cite{3}.

\textit{B. ZigBee:}
ZigBee standard is most commonly used standard. It has the capability to handle complex communication in low power communication devices (such as, nodes) with collision avoidance schemes. It consumes less power (nearly $60 mW$) and provides low data rate ($250 kbps$). Hardware support with encryption is featured by many ZigBee controllers to provide effective protection for communication in WBAN \cite{3}.


\textit{C. Medical Implant Communications Service (MICS):}
This band is specially designed for communication in WBAN. It is a short distance standard and is used to gather signals from different sensors on the body in a multi-hop structure. As compared to UWB, MICS has very low power radiation and thus is most suitable for the sensors used in UHC monitoring system \cite{3}.

\textit{D. IEEE 802.15.6 Ultra Wide Band (UWB):}
It provides very high bandwidth and data rate for communication.  It is used for localization of transmitters. When very high bandwidth is required in any application, UWB is the best choice. For example, when emergency or critical situation occurs, UWB with GPS (global positioning system) provides the best, short and traffic free route to the medical centre without any interference. User localization is usually important in hospitals or whenever, an emergency situation takes place. The advantage of UWB is that it is the only reliable method of localization. The drawback is receiver's complexity because of which it is not suitable for wearable applications in health monitoring \cite{3}.

\section{Wearable Sensors used for UHC}\label{sec:devices}
In this section, we briefly discuss wearable sensors used for UHC. Several types of tiny sensors are used in WBANs, which are attached to the body of a person to measure vital signs such as glucose level, Electrocardiogram (ECG), Electroencephalograph (EEG), detection of cancer cells etc., and surrounding parameters like temperature, atmospheric pressure, humidity etc. The size, shape and material of these sensors are of great importance. Moreover, these sensors must be compatible with the human body and their precise placement on the body, since these sensors are very sensitive and can harm human body. Therefore, these sensors are designed to be easily weared and provide comfort to patients. The sensors discussed in this section are normal clothing elements for the patients. Following is the discussion of sensors such as: wrist watch like an eWatch, wrist oximeter, chest belt, shoulder, necklace and wearable shirt type like a smart shirt/life shirt.

\textit{A. Wrist watch (eWatch):}
This device is just like a wrist watch, with a wrist body and a band attached to it. The major components of this device are: two Polydimethylsiloxane (PDMS) electrodes for ECG, a ribbon type temperature sensor, reflective flat type Pulse Oximeter Oxygen Saturation (SpO2) sensor, three printed circuit boards for analog and digital circuitry and other additional sensors. The device has a size of $60\times65\times15mm$ and weighs about $160g$ including one lithium polymer battery. Simple software is developed for this device to facilitate the users which are mostly elderly people. It consumes very low power since it is designed to be of very small in size.

\textit{B. Oximeter:}
Wrist pulse oximeter is a device that continuously monitors the patients pulse. It is small in size, light in weight, attached to the wrist and the sensor is placed on the finger. It is attached on the wrist using probe, which contains Light Emitting Diodes (LEDs) and photodiode to measure SpO2 Value. The wrist oximeter performs data acquisition $5 bytes/sec$ containing Photoplethysmographic (PPG) data sampled at $75 Hz$ and SpO2 values \cite{9}.

\textit{C. Chest belt:}
A wearable chest sensor belt has following essential components; main body of the belt, conductive fabric electrodes, an ECG sensor, an accelerometer sensor, double-layer PCB board, a wearable USN node, here, a conductive fabric electrode is used for obtaining ECG signals from the body.

\textit{D. Wearable shirt type (smart shirt/life shirt):}
Wearable shirt is also known as smart shirt/life shirt and it is another useful device for measuring the physiological parameters as well as physical activities help to improve patient's diagnosis. It is such a comfortable device that a patient does not feel the presence of any sensor or other components in the shirt. The device ensures a wide range of mobility. The wearable sensor nodes are designed to be tiny in size due to which they have limited power and computing capability. The designed wearable sensor node for
UHC features an ultra low power Texas Instrument MSP430 micro-controller with $10KB$ RAM, $48KB$ flash memory and $12-bit$ A/D converter \cite{11}.

\section{General WBAN Architectures}\label{sec:arch}

WBAN is used in health monitoring. The system discussed in \cite{1}, consists of three tiers: 1) TIER1 WBASN, 2) TIER2 Personal Server (PS), 3) TIER3 Medical Server (MS), as shown in Fig. 1.

\textit{A. Tier1 WBASN:}
It is the most predominant part of telemedical system and comprises of many intelligent nodes. Each node has the ability to sense, sample, process and
communicate different physiological signals. For example, heart activity is monitored by ECG sensor, muscle activity is monitored by EMG sensor, brain electrical activity is monitored by EEG sensor, blood pressure is monitored by a blood pressure sensor, while differentiation of the user's status and estimation of his/her activity level is done by motion sensors. PS sends initialization commands to each sensor and it also responds to queries from server. WBAN nodes must be able to meet the requirements for low power consumption. Because this ability is necessary to enable and prolong continuous monitoring, small size, minimum weight, consistent integration into a WBAN, standards based interface protocols, and calibration and customization specific to patients.

Sensor nodes can be patched to the clothes or shoes to constantly collect the information, store them locally and eventually transmit the
information to the PS after necessary processing. In case of an indication of emergency or critical situation during process of local analysis of the data, PS can send request for transmission of raw signals to MS. In short, each sensor node gets initialization command from PS and also responds to its queries. In order to ensure confidentiality of patient's information, data transfer at all tiers in UHC system must be encrypted. However, in critical cases, PS can also get direct connection with emergency medical services if the user desires to use this service.

\begin{figure}[h]
\begin{center}
\includegraphics[scale=0.40]{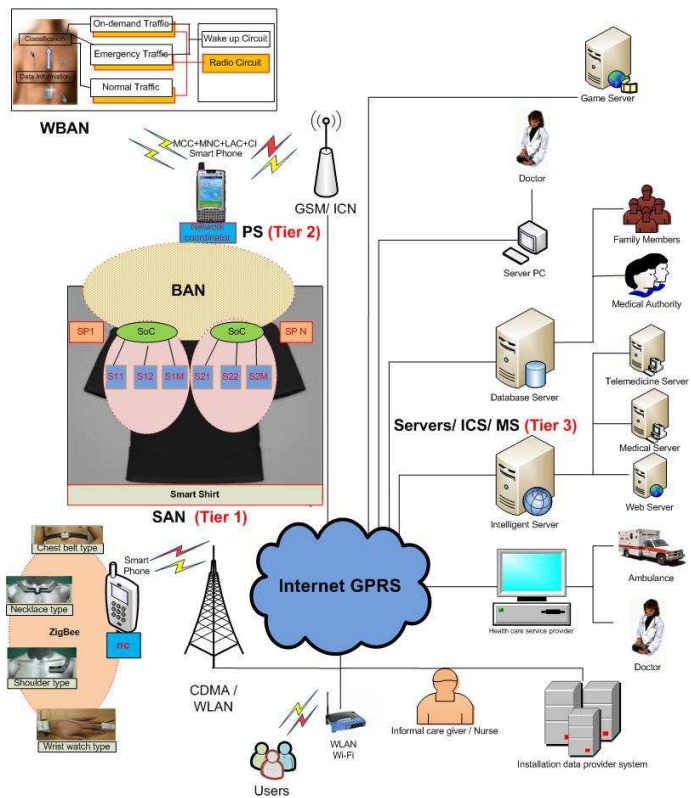}
\caption{Ubiquitous health care architecture}
\end{center}
\end{figure}

\textit{B. TIER2 PS:}
PS takes the information from sensor nodes about health status and transfer it to MS through WLAN or any internet service. A PS can be a laptop computer, handheld pocket PC or home PC, PDA, cell phone etc. WBAN nodes are interfaced by a PS through a network coordinator, which is responsible for implementing ZigBee (802.15.4) or Bluetooth connectivity. PS is ideal for elderly patients, who are home-bound, and it is used for communicating with MS. Interface of PS to WBAN comprises of a network configuration and management that includes various tasks. These tasks are node registration (e.g., type and number of sensors), initialization (e.g., state sampling frequency and operation mode), customization (e.g., run user specific adjustment), and setup of a secure communication (key exchange). On successful configuration of WBAN, network is handled by PS, which also ensures time synchronization, channel sharing, data recovery, data processing, and coalition of the data. PS is also responsible for patient's authentication information. To interface with MS, PS is configured with MS IP address. In order to ensure complete user mobility with secure and near real time health information provisioning, PS functions in such a manner that the information/data of the patient is transmitted when the link with MS is available.  However, when link is not available, PS stores the data locally and transmits it after the availability of communication channel or link.

\textit{C. TIER3 MS:}
The telemedical system is spread over a network that comprises individual monitoring systems connected to MS tier through internet. MS is optimally used for provision of services to a large number of individual users. Besides, this tier also provides service to a complex network comprising of interconnected services, medical personnel and HealthCare professionals. Keeping the records of registered users, variations in their health status and informs medical caregivers in case of emergency are main functions of MS. MS also forwards the instructions from physician regarding prescribed medicines and exercises to the users. The physician in this way, can monitor patient's vital information while sitting in his/her office through internet and ensures that patient follows the prescribed medicine and exercise.

\textbf{A server agent}:
Uploaded data and patient's record are examined by a server agent. It also creates an alarm in case of any critical medical situation. Large amount of data deposited through these services can be used for creating awareness among the patients.

%

\textbf{Multi-Tiers Advantages Over Typical Present Day Solutions}:
Multi-tier architecture provides a good solution for HealthCare problems in today's world. This section discusses use of multi-tier architecture and its advantages.

According to present day solution using multi-tier architecture, John is recovering from a heart attack. A prescribed rehabilitation scheme was suggested by the physician(s) for John, after he got discharged from the hospital. This scheme was to be adopted by John in home. By himself, he does not follow the exercise exactly as prescribed by the doctors. As a result, John's health recovery is much slower than what it should be in an expected time. The doctors are concerned about John's health and they do not have any other way to find out John's involvement for rehabilitation scheme.

Multi-tier health monitoring can provide good solution to John's rehabilitation scheme at low cost and minimum time. In this way, many tiny sensors are attached to John's body to monitor vital signs and assist John's exercise. Sensors or electrodes on the chest monitor the heart electrical activity while tiny inertial sensors measure John's body posture, movement, vital signs and intensity level of John's exercise. The information from the sensors can be gathered by physicians through internet for look up and depending on the information; they can prescribe new medicines and can adjust data threshold values. Any variation or change in the collected data is a measure that John is not following the doctor's prescription. This system thus reduces the cost, time and office visits to the physician. In case of a major emergency, PS can directly contact with the medical health staff, if the user pays for this service.

\textit{D. System Based Architecture with Physiological Signal Devices of UHC:}
UHC system architecture is shown in Fig. 1. This architecture consists of: Physiological Signal Devices (PSDs), a mobile system, a device provider system, a HealthCare service provider system, a physician system, a HealthCare personal system.

\textit{1) Physiological Signal Devices (PSDs):}
Physiological signals of the patient are measured by PSDs and then the data is transferred to mobile system using ZigBee 802.15.4.

\textit{2) Mobile System:}
It can display the physiological data from PSDs and with the help of Wireless Local Area Network (WLAN) or Code Division Multiple Access (CDMA) transfers the data to health care service provider system.

\textit{3) Device Provider System:}
Device installation data is supplied to the mobile system with the help of this system.

\textit{4) HealthCare Service Provider System:}
This system is like a portal where all the tasks regarding health care are performed.

\textit{5) Physician System:}
The patient's history is stored in health care service provider system which can be analyzed and scrutinized by the physician in this system.

\textit{6) HealthCare Personal System:}
The patient can examine and observe his own vital parameters and physiological signals after making a log in this system.

\textit{E. Integrated System Architecture of UHC Monitoring Systems:}
The most efficient architecture of UHC monitoring system is shown in Fig. 1, which consists of a hierarchical network of WBAN. The number of sensor platforms in WBAN, their packaging, size, shape, material and placement on human body depends on precise arrangement of integrated sensor nodes. The number of sensor nodes/platforms also depends on type of application being used in WBAN. The number of nodes and gyroscope sensors which are small in size. Accelerometer sensors are used to measure patient's position, posture and movement while gyroscope measures the level of patient's activity (e.g. running fast, walking, slow walking). WBAN consists of tiny sensors which are either planted or attached to the human body as patches or knitted to the clothes or implanted below the skin. The system architecture contains several networks that are hierarchically organized. This system is discussed as follows:

\textit{1) Sensor Area Network (SAN):}
Several sensors, \emph{Sij}, are combined into a single sensor platform, \emph{SPi}, with the help of wired or wireless interface. The sensors can be wired to \emph{SPi} or located on the sensor platform itself. Sensor platform can be seen in Fig. 1. Combination of two sensors; accelerometer and gyroscope is used by motion sensor platform for wearable monitoring applications.

\textit{2) Body Area Network (BAN):}
BAN integrates sensor platforms \emph{SPi} into a single monitoring system that is controlled by PS. Whereas, PS acts as a network
coordinator that gathers the information from all sensors and transfers it to BST. The communication in BAN can be wired in clothing or
implemented using short range wireless communication standards such as Bluetooth, Zigbee, UWB and MICS \cite{3} etc. Integrating information from individual sensors to users by home/gateways is used in WBAN systems to assist the medical staff in the hospitals.

\textit{3) Wide Area Network (WAN)}
Integration of multiple monitoring systems into a mobile health system through a cellular network is done by WAN. For connectivity, ubiquitous monitoring systems count on WAN using LANs such as WiFi when accessible.
\begin{itemize}
\item {\textbf{Server cloud}}:
Multiple servers such as MS containing patient's medical history, information servers, social network servers and other servers are incorporated by server cloud.
\end{itemize}

\textit{F) Traffic Based Architecture of WBAN for UHC Monitoring:}
WBAN traffic based architecture for UHC monitoring is classified into three types; (1) On-demand Traffic, (2) Emergency Traffic, (3) Normal Traffic.

\textit{1) On-demand Traffic:}
This type of traffic is activated by the doctor or the physician whenever, he/she needs any information for diagnostic recommendation. This traffic is further divided into following forms:
\begin{itemize}
\item {\textbf{Continuous}}:
In case of surgical events, On-demand traffic is continuous.
\end{itemize}

\begin{itemize}
\item {\textbf{Discontinuous}}:
When casual information is required, On-demand traffic is discontinuous.
\end{itemize}

\textit{2) Emergency Traffic:}
When WBAN nodes get beyond a predefined threshold, emergency traffic is initiated. Priority must be given to this type of traffic and it must be accommodated in less than one second. This type of traffic is not generated at regular intervals and is initiated in case of emergency or critical health condition.

\textit{3) Normal Traffic:}
As its name signifies, this kind of traffic takes place during normal condition when there is no emergency, time criticality and any event warranting on-demand traffic. It is basically a routine traffic that contains unnoticeable and routine health monitoring information about the patient and treatment of many diseases like gastrointestinal tract, cancer detection, neurological disorders and heart disease in normal time. This kind of normal data is collected and processed by the Network Coordinator (NC) which is activated by a radio circuit as per the application requirements. The wake up radio circuit has the ability to quickly respond to life critical events. With the objective of obtaining concerned recommendations, the NC is also linked to telemedicine and MSs. In short, a well integrated WBAN in the health care system caters for not only avoiding the occurrence of myocardial infarction and other life threatening diseases. Besides, it is also very handy for non-medical applications i.e., gaming and entertaining applications etc.

\textit{G. Components Based System Architecture:}
This architecture of UHC, as shown in Fig. 1, comprises of three components which are discussed as under:

\textit{1) Wearable Wireless Body Area Network (WWBAN):}
In this type of network, sensor(s) are attached to patient's body which collect and transmit suitable data to Intelligent Central Node (ICN) through Bluetooth communication protocol. Adoption of centralized architecture of star topology in WWBAN contributes to concentration of system intelligence on a central node, as shown in Fig. 1.

\textit{2) Intelligent Central Node:}
ICN is nothing but a smart phone with operating system that remains in communication with ICS through GPRS technology. ICN is connected to ICS and it performs the function of collecting and processing the data generated from different nodes in WWBAN. The collected data includes information pertaining to location area identifications such as: Mobile Country Code (MCC), Mobile Network Code (MNC), Location Area Code (LAC) and Cell Identification (CI) from GSM network. This information helps in determining the location of elderly patients with the help ICS. In the event of a change in collected parameters/information, ICN uses a comparison algorithm to decide whether to send information to ICS prior to transmit data from the sensors.


\textit{3) Intelligent Central Server:}
The purpose of ICS is to receive sensor data from all ICNs. The data received by the server is stored in database such that it can be analyzed independently without any human intervention. The latest uploaded patient's information on the server is compared with the already existing information to cure disease. In addition, it also provides information regarding prescription/recommendation of doctor or health care professional.


\textit{H. Wearable Smart Shirt Based Architecture for UHC and Activity Monitoring:}

\textit{1) Wearable Smart Shirt System:}
As shown in Fig. 1, this system contains a shirt with integrated WSNs, a BST and a server PC for distant monitoring. Smart Shirt is used to provide the individual physiological data. Then this data is transmitted in ad-hoc wireless communication for further processing using a wireless link. This is possible because Smart Shirt is compatible with wireless sensor network.

\textit{2) Wearable Sensor Node:}
The functionality of wearable sensor nodes is to get physiological data from human body and forward it to BST which in turn transmits data through a wireless link to the medical staff. A sensor node is small in size and consumes less power. It has low computation and communication capability and provides longer battery life for WBAN. To minimize the size of the sensor nodes, a board known as Universal Serial Bus (USB) programming board is designed as a separate module. This module is required only when the nodes are connected to a server PC for downloading an application or when the node itself acts as BST for data transmission.

\textit{3) Sensor Board}
The wearable sensor devices board which connects different sensors for various applications, is small in size, consumes low power and provide longer battery life.  In ECG, different electrodes are attached to the human body which measure heart's electrical signal and record potential difference between them. Three axis- accelerometer is a device which is used to monitor patient's behavior and physical activity in daily life. This is done with the help of fall detection system. Accelerometer signals can be collected by fall detection system and system determines that whether person has fallen or not.

\textit{4) Architecture of Wearable Smart Shirt with Integrated Sensors}
The measured accelerometer data and ECG data is transmitted to PS in WSN. ECG is used to monitor the heart status of the patient while an accelerometer is used to measure his physical activity. If both signals and data can be measured simultaneously, the determination of patient's disease can be improved.

To reduce the size of the wearable sensor node, a structure of two PCB stories is used, which comprises a wireless sensor node plate for communication with WSN and a sensor board plate with ECG interface and accelerometer. In this architecture, to reduce the size of sensor node, two round shape PCB boards are combined; one PCB board is wireless sensor node and second is sensor board.

Wireless sensor node is placed at the top position of two PCB stories, and the sensor board with ECG interface and accelerometer is placed at bottom position. This two stories structure of the sensor node reduces the wideness of a normal single story structure with sensors and provides convenience to wear it with two AAA size batteries. Two conductive electrodes knitted to smart shirt are extended from interface circuit of sensor board to get physiological ECG data from Smart Shirt.
%

\section{Path Loss in WBAN}\label{sec:pathloss}
WBAN is greatly influenced by the amount of path loss that occurs due to different impairments. Devices for WBAN are generally placed inside or on the body surface, therefore, losses between these devices would affect the communication and can degrade the performance monitoring in UHC. In the following sections, we study in detail about WBAN communication and path loss that occurs in it and how it affects the performance of UHC.

Reduction in power density of an electromagnetic wave introduces path loss \cite{4}\cite{12}. Path loss is mainly caused by free space impairments of propagating signal like refraction, attenuation, absorption and reflection etc. It also depends on the distance between transmitter and receiver antennas, the height and location of the antennas, propagation medium such as moist or dry air etc, and environment around the antennas like rural and urban etc \cite{12}. Path loss for WBAN is different from traditional wireless communication because it depends on both distance and frequency. Frequency is catered because body tissues are greatly affected by the frequency on which sensor device is working.

The path loss model in $dB$ between the transmitting and the receiving antennas as a function of the distance $d$ is computed by \cite{7}\cite{13} as:

\begin{equation}
PL(d)=PL(d_{o})+10nlog_{10}(\frac{d}{d_{o}}) + \sigma_{s}
\end{equation}\newline where, $PL(d_{o})$ is the path loss at a reference distance $d_{o}$, $n$ is the path loss exponent, and $\sigma_{s}$ is the standard deviation.

Path loss in WBAN is of great importance. UHC in WBAN works well when the path loss between the transmitter and receiver is at its minimum. Path loss in WBAN occurs due to many factors such as reflection, diffraction and refraction etc, from the body parts which may distort the signal and can cause interference at receiver located at a distant location. So, data may face distortion due to path loss which causes difficulty for medical team located at far distance to correctly retrieve data. Path loss in UHC will decrease the efficiency of monitoring different vital signs in human body at patient's level as well as at medical team's level. The main focus of this section is to minimize the path loss that occurs at different stages in WBAN. This increases the efficiency of UHC monitoring in BAN which is our main goal.

Path loss depends on distance as well as frequency and it is given in Eq. (2) and Eq. (3)

\begin{equation}
PL=20\log _{\left (10  \right )}\left ( \frac{4\pi d}{\lambda } \right )
\end{equation}
where, $L$ is the path loss in decibels, ${\lambda }$ denotes wavelength and $d$ specifies distance between transmitter and receiver \cite{12}.

As we know that: $\lambda=\frac{c}{f}$; however, the above Eq., can be rewritten as under:

\begin{equation}
PL=20\log _{\left (10  \right )}\left ( \frac{4\pi d f}{c } \right )
\end{equation}

\textit{A. Wireless Body Area Network}
In development of WBAN, one of the consideration is the characterization of the electromagnetic wave propagation from devices embedded inside the human body or close to it. A simple path loss model for WBAN is difficult to be driven in view of complex nature of human tissue structure and body shape. Since the antennas for WBAN application need to be placed on or inside the body, channel model has to take into consideration the effect of the body on radio propagation. To calculate the path loss in WBAN, three types of nodes are defined as under:\\
(a) \textbf{Implant node}: This type of node is embedded inside the body either below the skin or deeper.\\
(b) \textbf{Body Surface node}: This type of node is placed on the surface of human skin or maximum $2cm$ away.\\
(c) \textbf{External node}: This type of node is kept away from the body by a few centimeters upto a maximum of $5 meters$.\\
For body surface communication, it is also important to consider the distance between the transmitter and receiver around the body. If these are not placed on the same side in a straight line, then it allows the creeping wave diffraction to be also taken into account. As mentioned earlier, for external node communication, the distance between transmitter and receiver from the body vicinity is normally $3 meters$ away. However, in some cases, the maximum range of medical device can go upto 5 meters \cite{14}.

\textit{1) Effect of WBAN Antennas:}
In case of antennas placed on the surface or inside the body, it is influenced by its surroundings. It is therefore essential to understand the changes in the antenna patterns and other characteristics must also be taken into account in the scenarios requiring propagation measurements It is noticeable that the form factor of antenna is dependent on the requirements of applications. Different types of antennas are suitable for different applications e.g., for MICS, a circular antenna is used for pacemaker implant, while a helix antenna is most appropriate for a heart stent or urinary implant. Antenna's performance is greatly influenced by the form factor, which in turn effects the overall system performance. Antennas which take into account the characteristics of human body (such as, change in body tissues etc), are designed for measurements of channel model \cite{15}. Antennas used in WBAN communication are categorized into following two types \cite{16}:

(a) \textbf{Electrical antennas, such as dipole}:
Electrical antennas are generally used for On-Body communications. They are avoided for In-Body communications because electrical activity of these antennas is harmful for tissues and muscles of body. On-Body communications, through these antennas do not make any direct contact with the body tissues and muscles. Thus, not resulting in the heating of tissues.\\
(b) \textbf{Magnetic antennas, such as loop}:
Magnetic antennas are mostly used for In-Body and Implant communications. Magnetic antennas do not overheat the body tissues and is not dangerous to human body unlike electrical antennas. A loop of magnetic field is formed in magnetic antennas, which is within the defined range of the antenna, thus, these can communicate within this range not interfering with the body.

\textit{2) Characteristics of Human Body:}
For wireless communication in WBAN, human body is not considered an ideal medium for the propagation of signal. Human body consists of materials which contain different dielectric constants, thickness and impedance which may not be ideal for communication. Depending on the frequency of operations, human body may encounter many impairments and losses such as absorption, attenuation and diffraction etc. Therefore, the characteristics of human body should be kept in mind before designing the path loss model for WBAN \cite{17}.

\section{Scenarios of Path Loss in WBAN}\label{sec:scenarios}
There are different scenarios of path loss which can take place in WBAN, since sensor nodes can be implanted inside the human body either planted on the surface of the body or atmost $2mm$ away from the body surface. Since path loss is dependent on distance as well as frequency, therefore, the variations of these parameters in these scenarios will affect the path loss model. The simulation study is performed in MATLAB. A detailed discussion of these simulations is given as under.

\textit{A. In-Body Communication}
In order to study propagation characteristics inside human body, simulations are carried out using a 3D visualization scheme \cite{5}. The reason of using this scheme is that the study of physical parameters and their measurements are not feasible inside the human body. The antenna used in this study is a multi-thread magnetic loop antenna \cite{18}. Like in \cite{18}, we consider four models for our simulation which include: 1) Deep Implant to On-Body, 2) Near Surface Implant to On-body, 3) Deep Implant to Implant, and 4) Near Surface Implant to Implant. The path loss formula is same as \cite{7}\cite{13} and is given in Eq. (1) with a reference distance $d_{o}=50mm$ and frequency of $402-405 Mhz$. The path loss exponent and standard deviation values for implant to body surface models are given in Table. 1.

\begin{table}[!h]
\begin {center}
\begin{tabular}{|m{3cm}|m{2cm}|m{1cm}|m{1cm}|}
\multicolumn{3}{c}{\hspace{-3 cm}\textbf{Table. 2. Implant to Body Surface}}\\
\hline
{\textbf{Models}} & \textbf{Path Loss in dB} & \textbf{n} & $\boldsymbol{\sigma_{s}(dB)}$   \\
\hline
Deep Tissue Implant to Body Surface & 46.14 & 4.86 & 7.25\\
\hline
Near Surface Implant to Body Surface & 47.81 & 4.532 & 6.23\\
\hline
\end{tabular}
\end{center}
\end{table}

From Table. 2 we conclude that the path loss at a reference distance for deep tissue implant to body surface is less than that of near surface implant to body surface because of high distance. The path loss exponent and standard deviation values for implant to implant models is given in Table 2.

\begin{table}[!h]
\begin{center}
\begin{tabular}{|m{3cm}|m{2cm}|m{1cm}|m{1cm}|}
\multicolumn{3}{c}{\hspace{-4 cm}\textbf{Table. 3. Implant to Implant}}\\
\hline
\textbf{Models} & \textbf{Path Loss in dB} & \textbf{n} &  $\boldsymbol{\sigma_{s}(dB)} $  \\
\hline

Deep Tissue Implant to Implant & 35.55 & 5.71 & 8.36\\

\hline

Near Surface Implant to Implant & 41.25 & 5.12 & 8.95\\
\hline

\end{tabular}
\end{center}
\end{table}

\begin{figure*}[!t]
  \centering
   \subfigure[Deep Tissue Implant to Body surface Path Loss]{\includegraphics[height=4 cm,width=4.7 cm]{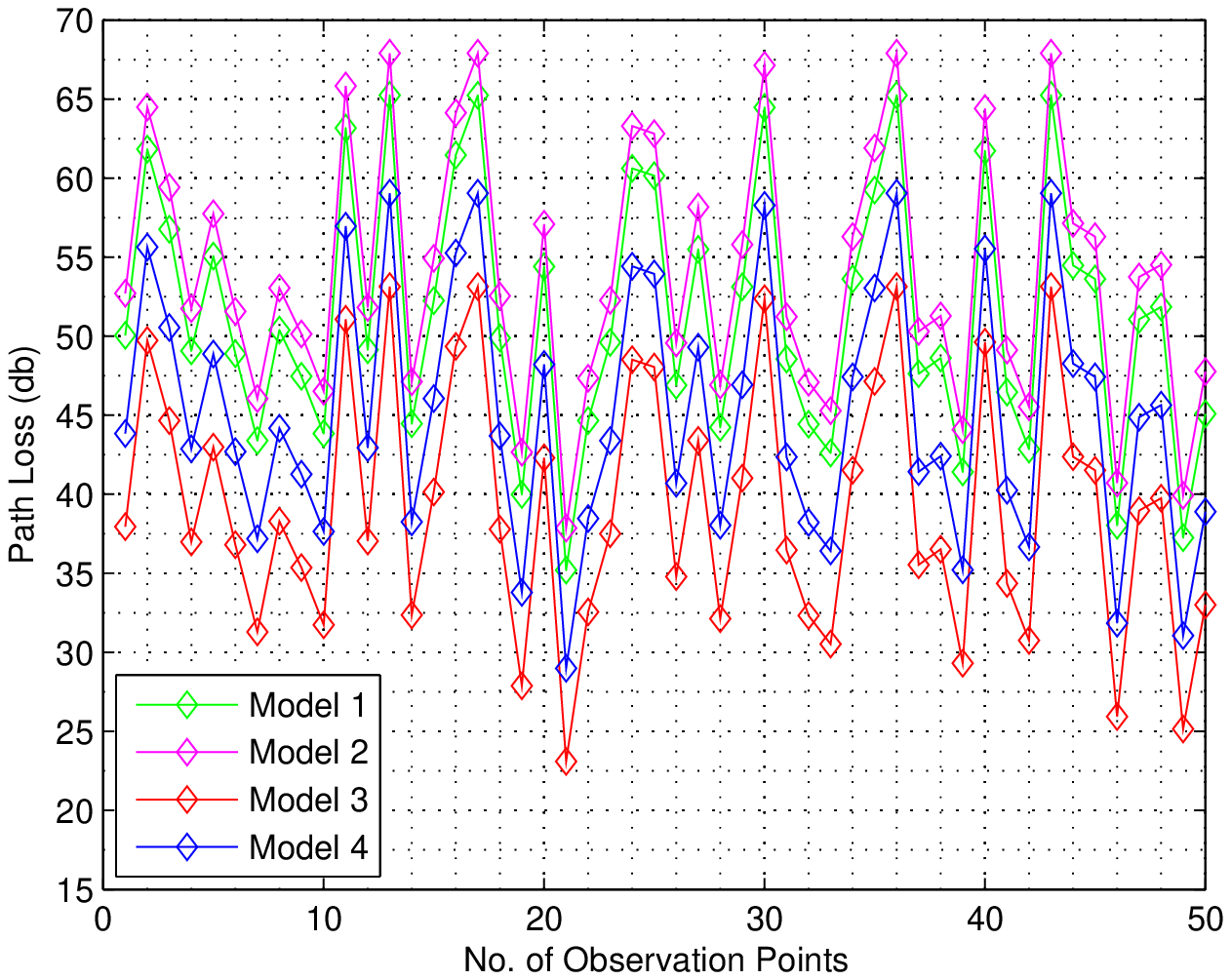}}
   \subfigure[Amplitude Attenuation in On-Body]{\includegraphics[height=4 cm,width=4.7  cm]{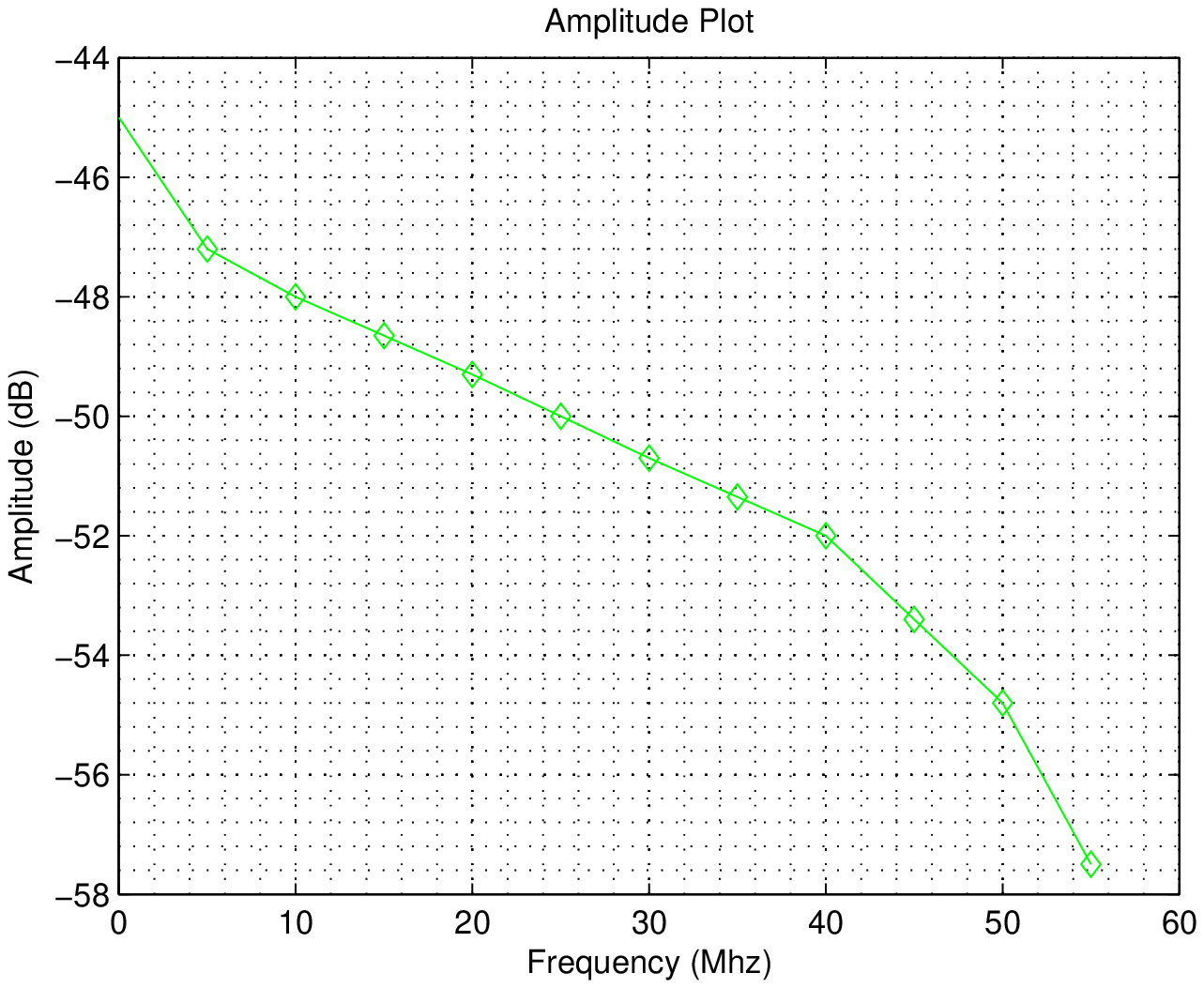}}
   \subfigure[Phase Distortion in On-Body]{\includegraphics[height=4 cm,width=4.7  cm]{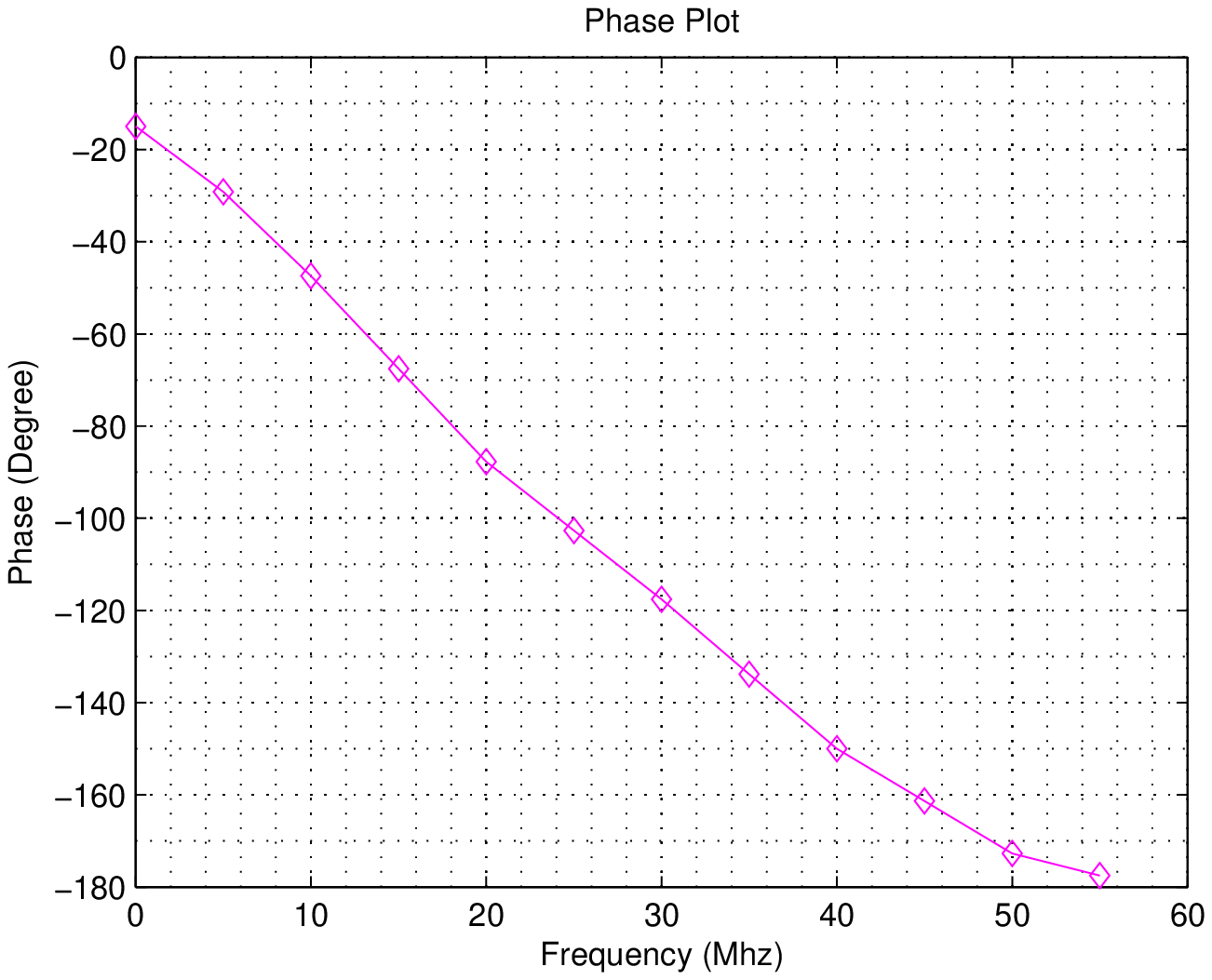}}
   \subfigure[Channel Output for On-Body Communication]{\includegraphics[height=4 cm,width=4.7  cm]{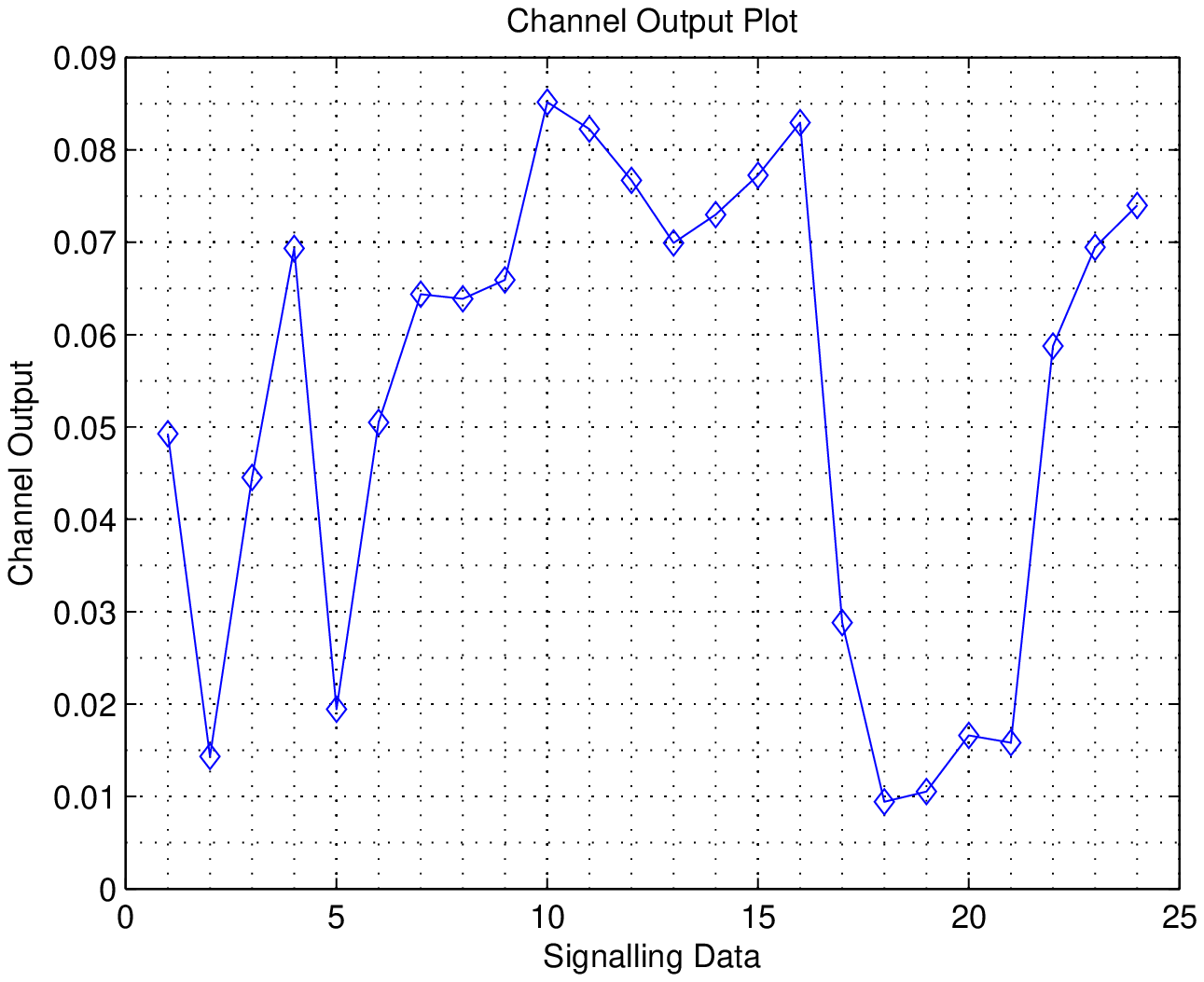}}
   \subfigure[Path Loss vs Distance for On-Body Communication]{\includegraphics[height=4 cm,width=4.7  cm]{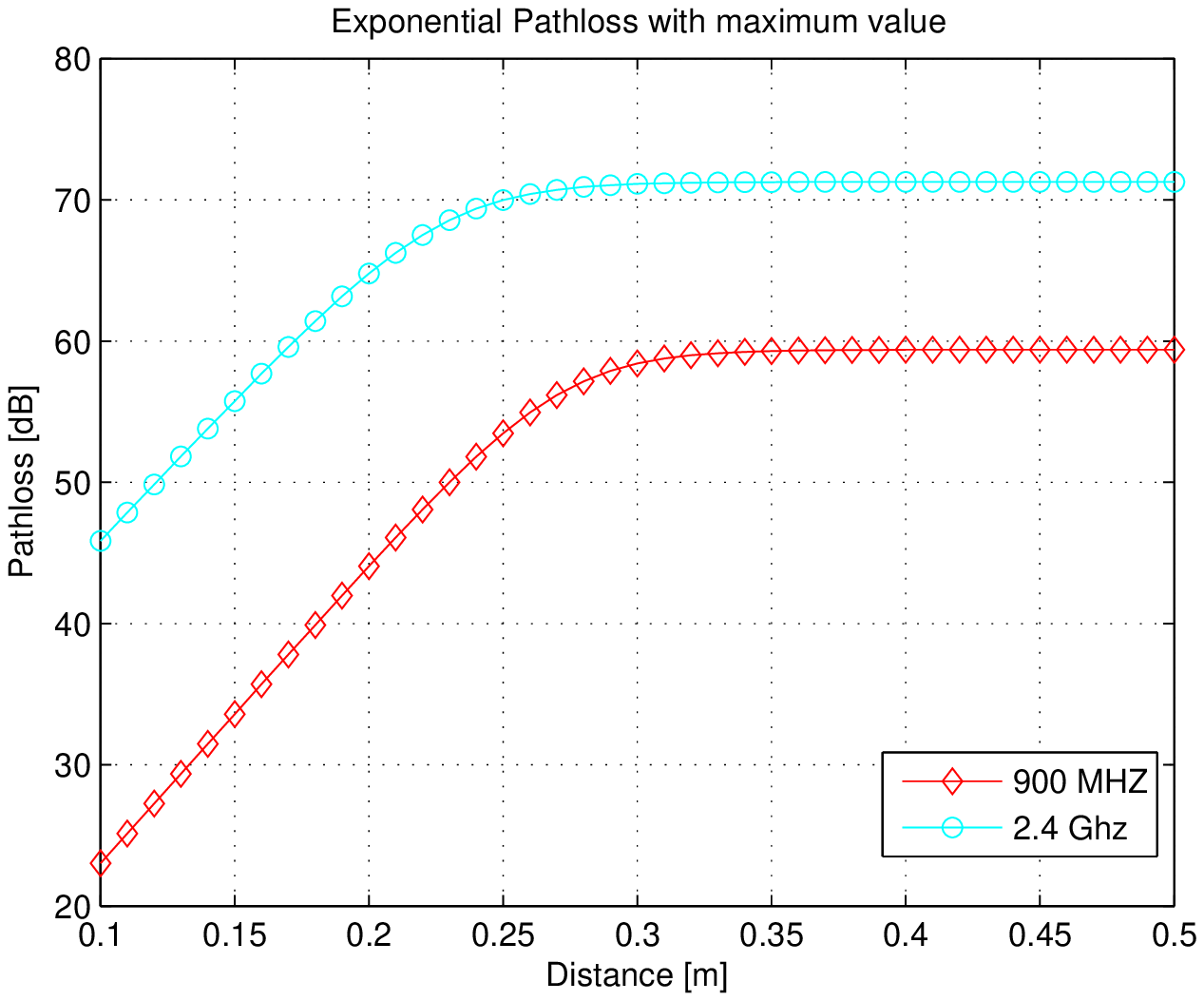}}
  \caption{Analytical Results}
\end{figure*}

%

Figure. 2(a) describes the simulation results of path loss from deep tissue implant node to body surface node communication (Model 1). The simulation is carried out between no. of observation points between the deep tissue implant node and reference node placed at some distance along with path loss at each point of observation. The graph shows fluctuations in path loss at each observation point. The no. of observations are fixed at $50$ for each model in the simulation. Deep Tissue Implant to Body Surface (Model 1) curve in Fig. 2(a) is taken as reference for other three models. Since, no. of observation points are fixed at $50$, therefore, the fluctuations are same for all models. However, increase or decrease in the path loss value is dependent upon the model which is being used. For near surface implant to body surface path loss model (Model 2), there is an increase of $4dB$ in path loss at each observation point from the reference model. For deep implant to implant and near surface implant to implant models, increase of $15db$ and $9dB$ in path loss, respectively is noticed in Fig. 2(a). For deep tissue implant to implant path loss model (Model 3), a decrease of $11dB$ in path loss occurs at each observation point from our reference model (Model 1). The decrease is evident because distance between the nodes are smaller in this model (i.e., Model 3) from the reference model (Model 1). With comparison to near surface implant to body surface path loss model (Model 2), a decrease of $15dB$ in path loss is observed. This is because of further lessening of distance between the nodes, and decrease of $6dB$ in path loss from near surface implant to implant model (Model 3). For near surface implant to implant path loss model (Model 4), a decrease of $5dB$ in path loss is obtained at each of the observation points from reference model (Model 1), $9dB$ for near surface implant to body surface path loss model (Model 2), while an increase of $6dB$ in path loss for deep tissue implant to implant path loss model (Model 3). If we further increase the no. of observation points, the fluctuations in path loss will be more sudden. This is due to different impairment factors such as refraction, diffraction, reflection etc.

\textit{2) On-Body Communication}
For On-Body communications in WBAN, placement of sensors and actuators on the body surface is of great importance. Simple path loss model that takes into account the placement of sensors on the body and their communication with respect to body postures and movements is required. Channel response output of the On-Body communication as well as the frequency response can be easily found out. UHC monitoring in WBANs depends on both In-Body and On-Body communications of sensor nodes \cite{6}.
%

Amplitude attenuation of the signal with respect to frequency for On-Body communication is depicted in Fig. 2(b). As frequency increases, attenuation of amplitude also increases since the channel undergoes impairments. Thus, these impairments degrades the intensity of signal, as it travels from transmitter to the receiver node planted on the human body.


Phase distortion of the signal with respect operating frequency for On-Body communication is obvious from Fig. 2(c). The direct relationship exists here as well; with the increase in frequency the phase distortion of the signal increases in a linear fashion and vice versa. Each component of the signal is distorted in phase and if relationship is not linear then there will be different phase distortion at different frequencies. From UHC point of view, both amplitude attenuation and phase distortion of the signal for On-Body communication should be eradicated to achieve better monitoring results.



Figure. 2(d) describes the channel output of On-Body communication with respect to the signalling data. The signalling data is uni-polar Non Return to Ground (NRG) stream of ones and zeros, respectively. Depending on this data, the channel output fluctuates with it having a higher output when the signalling data stream of more ones than zeros exists and lower output when the signalling data stream consists of more zeros than ones.


As, discussed earlier that path loss depends on the distance between the transmitting and receiving antenna/node as well as frequency of operation. The simulation results shown in Fig. 2(e) are carried out for two different frequencies (i.e., $900 Mhz$ and $2.4 Ghz$), as, it is obvious from Eq. (5) of path loss model. Since, direct relationship exists between path loss and distance between transmitter and receiver, therefore, by increasing distance, path loss increases linearly. Also, path loss has a direct relationship with frequency, thus, at $2.4 Ghz$ the path loss curve is  higher then that at $900 Mhz$.

\section{Conclusion}\label{sec:conclusion}
WBAN is an emerging domain in the field of wireless communication. It comprises of many tiny sensors placed on or inside the body. These sensors measure patient's vital information and transfer it to medical personnel for diagnosis. WBAN has many applications, most important of which is in UHC. With UHC, patients are not required to visit doctor frequently. They can get diagnosis of their disease while sitting at home. Nowadays, a lot of work is going on to make low power sensors and devices that can be used in UHC. In this paper, an analytical survey is done on different architectures of WBAN used in UHC. These architectures are suited for different applications. Wearable devices and standards used in WBAN are also discussed. Different standards are used depending on the type of application. Path loss in WBAN and its effects are also discussed in detail. Simulations for In-Body and On-Body communication are also performed. The results for On-Body communications show that path loss increases between transmitter and receiver with increase in distance and frequency. Similarly, phase distortion and attenuation also increases with frequency. Moreover, path loss in different models of In-Body communication is also carried out. Finally, the summarization of architectures and path loss in On-Body and In-Body communications is also presented.

\end{document}